\documentclass[]{spie}  

\usepackage{amsmath,amsfonts,amssymb}
\usepackage{graphicx}
\usepackage[colorlinks=true, allcolors=blue]{hyperref}

\usepackage{tabularx,booktabs}
\newcolumntype{Y}{>{\centering\arraybackslash}X}

\usepackage{fancyhdr}
\title{Bone Suppression on Chest Radiographs With Adversarial Learning}

\author[a]{Jia Liang*}
\author[a~]{Yu-Xing Tang*}
\author[a]{You-Bao Tang}
\author[b]{Jing Xiao}
\author[a]{Ronald M. Summers}
\affil[a]{Imaging Biomarkers and Computer-aided Diagnosis Laboratory, Radiology and Imaging Sciences, National Institutes of Health Clinical Center, Bethesda, USA}
\affil[b]{Ping An Technology Co., Ltd., Shenzhen, China}

\authorinfo{*Equal contribution. J. Liang is with Stanford University. Work done during his internship at NIH.\\
Further author information: (Send correspondence to Y. X. Tang and R. M. Summers)
\\E-mail: \{yuxing.tang, youbao.tang, rms\}@nih.gov}

\begin{document} 

\maketitle

\begin{abstract}
 Dual-energy (DE) chest radiography provides the capability of selectively imaging two clinically relevant materials, namely soft tissues, and osseous structures, to better characterize a wide variety of thoracic pathology and potentially improve diagnosis in posteroanterior (PA) chest radiographs. However, DE imaging requires specialized hardware and a higher radiation dose than conventional radiography, and motion artifacts sometimes happen due to involuntary patient motion. In this work, we learn the mapping between conventional radiographs and bone suppressed radiographs. Specifically, we propose to utilize two variations of generative adversarial networks (GANs) for image-to-image translation between conventional and bone suppressed radiographs obtained by DE imaging technique. We compare the effectiveness of training with patient-wisely paired and unpaired radiographs. Experiments show both training strategies yield ``radio-realistic'' radiographs with suppressed bony structures and few motion artifacts on a hold-out test set. While training with paired images yields slightly better performance than that of unpaired images when measuring with two objective image quality metrics, namely Structural Similarity Index (SSIM) and Peak Signal-to-Noise Ratio (PSNR), training with unpaired images demonstrates better generalization ability on unseen anteroposterior (AP) radiographs than paired training. 
\end{abstract}

\keywords{Generative Adversarial Networks, Bone Suppression, Dual-Energy Chest Radiography, Image-to-image Translation}

\section{Introduction}
\label{sec:summary} 


According to the Centers for Disease Control and Prevention, 50,000 people die from pneumonia in the United States every year~\footnote{\url{https://www.cdc.gov/dotw/pneumonia/index.html}}. Previous studies~\cite{macmahon2008dual, li2012improved, schalekamp2014bone, baltruschat2019does} suggested that dual-energy chest radiographs (chest X-ray or CXR) can improve diagnostic accuracy for finding abnormalities, especially focal pneumonia over standard chest radiography. Dual-energy (DE) chest X-rays separate images of bones and soft tissues by making use of the differential reduction of low-energy X-ray photons by calcium. However, the acquisition of dual-energy chest X-rays increases the radiation dose to the patients and requires special, expensive equipment. As a result, researchers have been exploring methods to obtain bone suppressed chest X-rays from standard chest X-rays, and substantial progress has been made~\cite{huo2014bone}. 

The bone suppression techniques could generally be categorized into deep learning and non-deep learning approaches. Non-deep learning approaches usually first locate the lung and ribs border and, then, use vertical intensity profiles to refine the final bone shadows~\cite{chen2014bone}.  As deep learning is further developed for chest radiograph analysis~\cite{Wang_CVPR2017, Tang_MLMI, tang19MIDL, tang2019deep}, the deep learning based bone suppression~\cite{oh2018learning, yang2017cascade, eslami2019image} also gradually gains more popularity since its greater power and flexibility to represent characteristics of different structures in the chest X-rays. One of the earlier deep learning approaches~\cite{chen2014bone} uses multiple massive-training artificial neural networks to first obtain a bone image from single energy chest X-ray. Then the bone image is subtracted from the single energy chest X-ray to obtain the virtual soft-tissue image. 

We propose to use generative adversarial networks (GANs)~\cite{Goodfellow_GAN} to learn bone suppression from dual-energy chest radiographs. GAN has gained much attention for its ability to generate realistic-looking synthetic images~\cite{tang2019ct, tang2019tuna}. GAN is composed of two networks, namely a generator and a discriminator. The generator creates images similar to the training set, and the discriminator tries to differentiate the true images from the training set and the fake images from the generator. When GAN is trained, the generator is able to generate images that are indistinguishable from the original training set. For our specific problem of bone suppression in the standard chest X-rays, the generator is able to learn a mapping from the standard chest X-rays to virtual soft-tissue images, by making use of dual-energy chest radiographs. In this work, we exploit two variations of GANs, namely, Pix2Pix~\cite{isola2017image} trained with patient-wisely paired radiographs and Cycle-GAN~\cite{zhu2017unpaired} trained with unpaired radiographs. Quantitative and qualitative experimental analysis verifies that image-to-image translation using adversarial learning is a feasible means to suppress bone structures and import minimal motion artifacts in standard radiographs. We also find that unpaired training using only posteroanterior (PA) chest radiographs yields better generalization ability on unseen anteroposterior (AP) radiographs.

\subsection{Purpose}
\label{sec:purpose}
To determine the feasibility and to compare the effectiveness of using variations of generative adversarial networks to suppress bones (e.g., ribs and clavicles) from standard frontal-view chest radiographs, by learning from paired or unpaired dual-energy chest radiographs.

\section{METHOD}
\begin{figure*}[t]
  \centering
  \includegraphics[width=\linewidth]{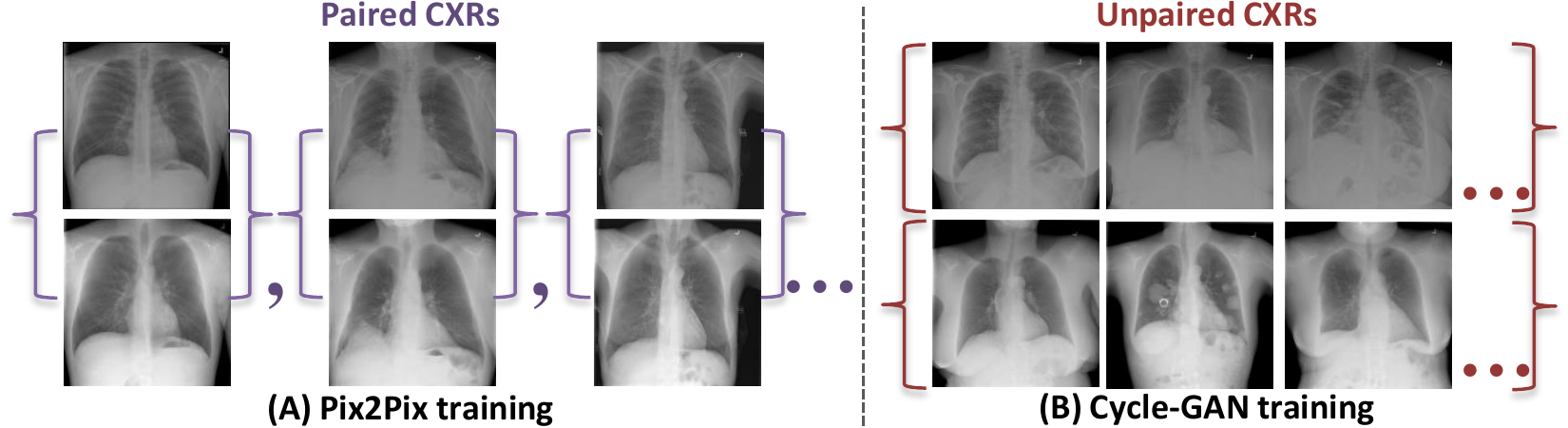}
  \caption{(A) Paired training radiographs consist of training pairs where the correspondence within each radiograph pair exist. (B) Unpaired training radiographs consist of a source set and a target set, with no information provided as to which standard CXR matches which bone suppressed CXR. \textbf{Top} row: standard chest radiographs. \textbf{Bottom} row: bone suppressed chest radiographs using the dual-energy technique.}
  \label{fig1}
\end{figure*}

Dual-energy subtraction imaging captures two or three radiographs of the same patient with different energy levels of X-ray exposures. One of the captured images highlights only the bones based on a specific energy level. Thus, the suppressed bone image can be estimated by combining the acquired standard chest X-ray image which includes both the soft tissue and bones and bone-only image. Therefore, we are motivated to utilize image-to-image translation techniques to translate a standard radiograph into a soft tissue only radiograph, thereby suppressing the bone structures. In this work, we adopt paired and unpaired training to accomplish this task (See Figure~\ref{fig1}).

\subsection{Framework for Paired Image-to-Image Translation}
We first adopt a variation of Pix2Pix~\cite{isola2017image} for the paired training of generative adversarial networks (GANs). The Pix2Pix model works by training on pairs of images, in this work, namely standard CXRs to bone suppressed CXRs, and then attempts to generate a corresponding output CXR without bones from a standard CXR. The Pix2Pix model is a type of conditional GAN, where the generation of the output CXR is conditional on an input CXR, in this case, a source image $X_s$. The discriminator $D$ sees both the source CXR and the generated target CXR $X_t$ and decides if it is a ground truth CXR or from the generator $G$. The generator tries to minimize the pairwise $L1$ distance and generate plausible bone suppressed CXRs to fool the discriminator. The adversarial loss function is:
\begin{equation}
\label{eq:gan_loss}
        \min_{G} \max_{D} \mathbb{E}_{X_s,X_t} 
        \Big[ \log D((X_s, G(X_s))) \Big]
        + \mathbb{E}_{X_s,X_t}
 		\Big[ \log \big ( 1 - D(G(X_s,X_t)) \big ) \Big].
\end{equation}
Please refer to Pix2Pix~\cite{isola2017image} for more details about the model training.

\subsection{Framework for Unpaired Image-to-Image Translation}
Pix2Pix works only when two image spaces are pre-formatted into a single $X_s/X_t$ image that held both tightly-correlated images. However, in clinical scenarios, there are far more standard, conventional radiographs than paired DE radiographs. In this case, we propose to use unpaired CXRs from source domain $S$ and target domain $T$ to train an unsupervised image-to-image translation model for bone suppression. Therefore, we adopt Cycle-GAN~\cite{zhu2017unpaired} for this task. A CycleGAN consists of two generators and two discriminators. The discriminators $D_S$ and $D_T$ classify an input CXR as real or fake. 
$D_T$ encourages the generator $G_{S\rightarrow{T}}$ to learn the mapping $S\rightarrow{T}$ and translate source CXRs $X_S$ into outputs indistinguishable from target domain $T$, and vice versa for $D_S$ and $G_{T\rightarrow{S}}$. In addition to adversarial losses, two cycle-consistency losses, namely forward cycle-consistency loss and backward cycle-consistency loss, are also used to regularize the model to ensure the transform from one domain to the other and back again to the original domain. The advantage of unpaired training is that the direct correspondence between individual CXRs is not required in two domains. Thus unpaired training might be more robust to unseen CXRs not closely aligned with the source distribution, e.g., anteroposterior CXRs unseen in the training. Please refer to Cycle-GAN~\cite{zhu2017unpaired} for more details about the loss functions and unpaired training. 

\begin{figure*}[t]
  \centering
  \includegraphics[width=0.95\linewidth]{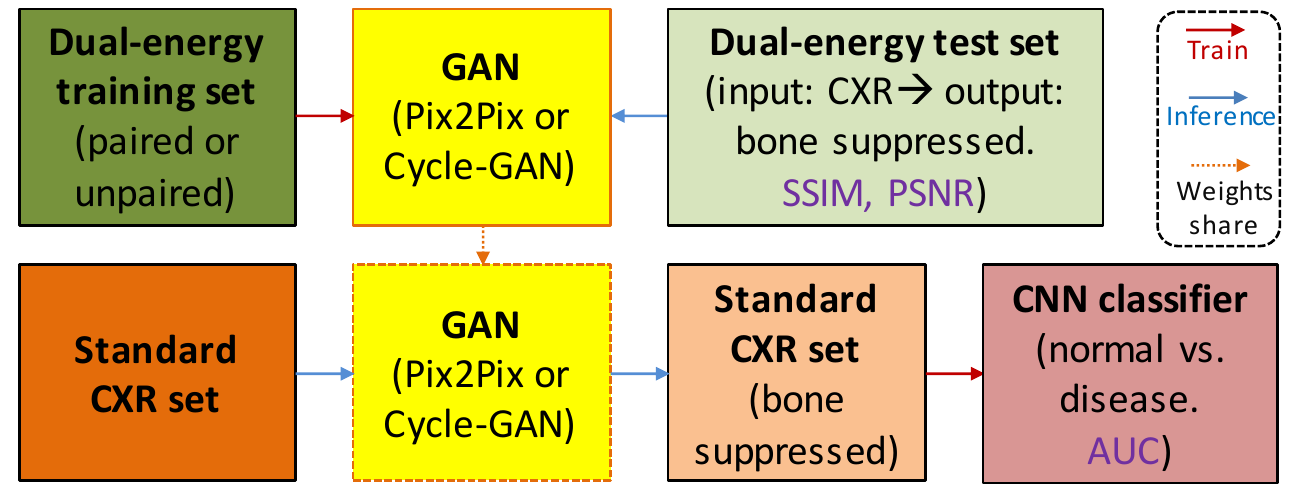}
  \caption{Work-flow of the bone suppression and evaluation framework.}
  \label{fig:wf}
\end{figure*}

\section{EXPERIMENTAL RESULTS}
\label{sec:result}

\subsection{Datasets}
\label{sec:data}
In this study, we experiment with two different datasets. First, we train and evaluate the image-to-image translation models for bone suppression on a dataset of 1,867 anonymized dual-energy PA chest radiographs collected from the picture archiving and communication system (PACS) of our institute. This dataset is randomly split to 7:1:2 for training, validation, and testing. We evaluate the models with two different objective image quality metrics, namely Structural Similarity Index (SSIM)~\cite{wang2004image} and Peak Signal-to-Noise Ratio (PSNR)~\cite{seshadrinathan2010study}. Then, we use the trained models on this dataset to generate bone suppressed radiographs on a subset~\footnote{\url{https://www.kaggle.com/c/rsna-pneumonia-detection-challenge/data}} of the NIH chest X-ray dataset~\cite{Wang_CVPR2017} with both PA and AP radiographs, containing 8,525 normal radiographs and 17,159 radiographs with abnormalities. Among these radiographs on the second dataset~\cite{shih2019augmenting}, 1,532 normal and 3,000 abnormal images are used to evaluate the binary classification performance, in terms of AUC (the area under the receiver operating characteristic curve), of the bone suppression models. The two datasets are denoted as ``dual-energy dataset'' and ``standard dataset'', respectively. We set the input and output image size to be 512 pixels as a trade-off between better image quality and affordable computational load for both paired and unpaired training and testing. The work-flow is shown in Figure~\ref{fig:wf}.
 
\subsection{Results}
\subsubsection{Quantitative Results}
We quantitatively evaluate the quality of generated bone suppressed radiographs by comparing them with the soft-tissue only images of the ``dual-energy'' test set, and then evaluate the normal versus abnormality binary classification results (using VGG-19~\cite{Simonyan15} as a classifier) on the ``standard'' test set using generated images. The results are shown in Table \ref{tab:results}. The framework trained with paired images slightly outperforms the unpaired counterpart on the dual-energy dataset. But when the framework is extended to the standard dataset which contains both PA and AP radiographs, the unpaired training shows better generalization ability, given the fact that the classification result is higher than the paired training. We could not evaluate the SSIM and PSNR on the standard dataset since ground-truth soft tissue images on this dataset are unavailable. 

\begin{table}[ht]
\caption{Comparison of paired and unpaired training on two different test sets. Higher scores are better (indicated in bold).} 
\label{tab:results}
\begin{center}
\begin{tabularx}{0.6\textwidth}{c *{3}{Y}}
  
\toprule
 & \multicolumn{2}{c}{Dual-energy}  
 & Standard\\
\cmidrule(lr){2-4} 
 Method & SSIM& PSNR & AUC\\
\midrule
 Paired  & \textbf{0.867$\pm$0.011} & \textbf{36.078$\pm$0.305} & 0.948$\pm$0.004\\
 Unpaired  &  0.855$\pm$0.010 & 34.820$\pm$0.293 & \textbf{0.953$\pm$0.003}\\
\bottomrule
\end{tabularx}
\end{center}
\end{table}

\subsubsection{Qualitative Results}
\begin{figure*}[t]
  \centering
  \includegraphics[width=\linewidth]{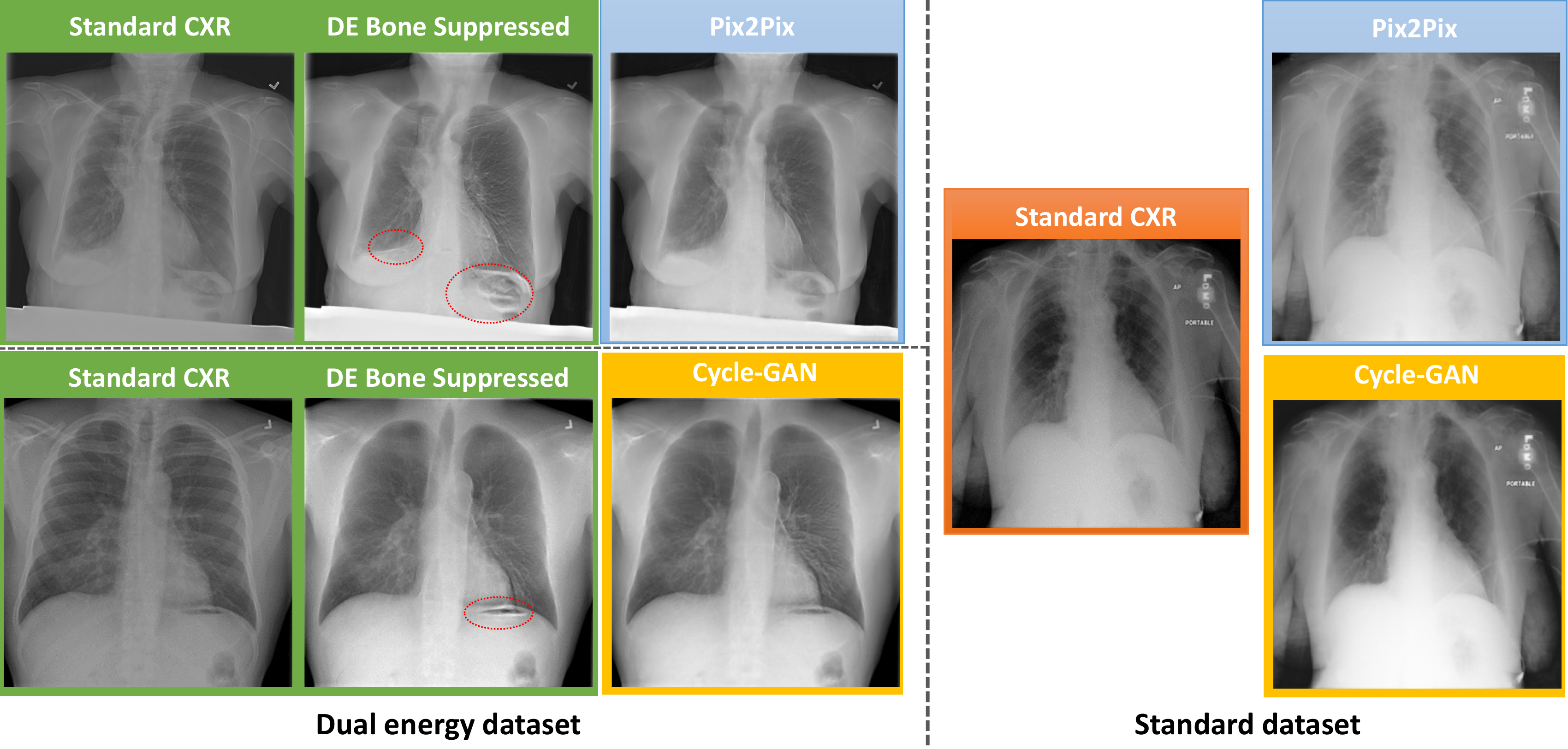}
  \caption{Some examples of real and corresponding generated bone suppressed chest X-rays. Left, PA chest X-rays from the dual-energy dataset. Top: Green area shows a pair of dual-energy CXR (a standard one and a bone-suppressed one as ground-truth), and blue shows bone suppressed image generated by the paired Pix2Pix model. Bottom: Green area shows another pair of dual-energy CXR, and golden yellow shows the suppression results by the unpaired Cycle-GAN model. Red ovals show motion artifacts by dual-energy subtraction technique. Right: An input AP chest X-ray from a standard dataset and outputs of bone suppressed CXRs from Pix2Pix and Cycle-GAN.}
  \label{fig:results}
\end{figure*}

We show some examples of bone suppressed CXRs generated by the paired and unpaired training frameworks on two different datasets in Figure \ref{fig:results}. As can be seen from the figure, both paired and unpaired training of GANs are able to generate bone suppressed CXRs of high quality for PA CXRs. We find that the GAN models introduce minimal motion artifacts compared with the dual-energy subtraction technique. The reason is that there is only a small portion of training data in the DE dataset contains motion artifacts. The image-to-image translation models tend to learn the majority of information from the entire data distribution. This characteristic of adversarial learning models can be considered as yet another main advantage of automatic bone suppression in addition to less radiation exposure. The visualized results on AP CXRs also showed the superiority of Cycle-GAN trained with unpaired CXRs over Pix2Pix trained with paired data. A possible reason is that Cycle-GAN is not strictly constrained by paired CXRs in the training, leading to better generalization to unseen AP radiographs than Pix2Pix.

\section{Conclusion}
We proposed to use generative adversarial networks to learn to suppress bone structures on chest radiographs. Experimental evaluations on two different NIH chest X-ray datasets validate the effectiveness of the framework on suppressing bones. The framework trained with unpaired posteroanterior radiographs generalized better to unseen anteroposterior radiographs, showing great potential to facilitate image interpretation in clinical scenarios where both PA and AP radiographs exist. As proof of concept,  we focused our evaluations on bone suppression. But this framework can be readily extended to a wider range of applications such as bone fracture or lesion detection on bone images generated using the adversarial learning method.

\section{Acknowledgments}
This research was supported by the Intramural Research Program of the National Institutes of Health Clinical Center and by the Ping An Technology Co., Ltd. through a Cooperative Research and Development Agreement. The authors thank NVIDIA for GPU donation.

\bibliography{report-full} 

\begin{thebibliography}{10}

\bibitem{macmahon2008dual}
MacMahon, H., Li, F., Engelmann, R., Roberts, R., and Armato, S., ``Dual energy
  subtraction and temporal subtraction chest radiography,'' {\em Journal of
  thoracic imaging}~{\bf 23}(2),  77--85 (2008).

\bibitem{li2012improved}
Li, F., Engelmann, R., Pesce, L., Armato, S.~G., and MacMahon, H., ``Improved
  detection of focal pneumonia by chest radiography with bone suppression
  imaging,'' {\em European radiology}~{\bf 22}(12),  2729--2735 (2012).

\bibitem{schalekamp2014bone}
Schalekamp, S., van Ginneken, B., van~den Berk, I.~A., Hartmann, I.~J.,
  Snoeren, M.~M., Odink, A.~E., van Lankeren, W., Pegge, S.~A., Schijf, L.~J.,
  Karssemeijer, N., et~al., ``Bone suppression increases the visibility of
  invasive pulmonary aspergillosis in chest radiographs,'' {\em PloS one}~{\bf
  9}(10),  e108551 (2014).

\bibitem{baltruschat2019does}
Baltruschat, I.~M., Steinmeister, L., Ittrich, H., Adam, G., Nickisch, H.,
  Saalbach, A., von Berg, J., Grass, M., and Knopp, T., ``When does bone
  suppression and lung field segmentation improve chest x-ray disease
  classification?,'' in [{\em 2019 IEEE 16th International Symposium on
  Biomedical Imaging (ISBI 2019)}{\nolinebreak\hspace{0.1em}]},   1362--1366,
  IEEE (2019).

\bibitem{huo2014bone}
Huo, Z., Xu, F., Zhang, J., Zhao, H., Hobbs, S.~K., Wandtke, J.~C., Sykes,
  A.-M., Paul, N., and Foos, D., ``Bone suppression technique for chest
  radiographs,'' in [{\em Medical Imaging 2014: Image Perception, Observer
  Performance, and Technology Assessment}{\nolinebreak\hspace{0.1em}]},   {\bf
  9037},  90370D, International Society for Optics and Photonics (2014).

\bibitem{chen2014bone}
Chen, S. and Suzuki, K., ``Bone suppression in chest radiographs by means of
  anatomically specific multiple massive-training anns combined with total
  variation minimization smoothing and consistency processing,'' in [{\em
  Computational Intelligence in Biomedical
  Imaging}{\nolinebreak\hspace{0.1em}]},   211--235, Springer (2014).

\bibitem{Wang_CVPR2017}
Wang, X., Peng, Y., Lu, L., Lu, Z., Bagheri, M., and Summers, R.~M.,
  ``Chest{X}-ray8: Hospital-scale chest x-ray database and benchmarks on
  weakly-supervised classification and localization of common thorax
  diseases,'' in [{\em Proceedings of the IEEE Conference on Computer Vision
  and Pattern Recognition}{\nolinebreak\hspace{0.1em}]},   2097--2106 (2017).

\bibitem{Tang_MLMI}
Tang, Y., Wang, X., Harrison, A.~P., Lu, L., Xiao, J., and Summers, R.~M.,
  ``Attention-guided curriculum learning for weakly supervised classification
  and localization of thoracic diseases on chest radiographs,'' in [{\em
  Machine Learning in Medical Imaging}{\nolinebreak\hspace{0.1em}]},
  249--258, Springer (2018).

\bibitem{tang19MIDL}
Tang, Y.-B., Tang, Y.-X., Xiao, J., and Summers, R.~M., ``{XLSor}: A robust and
  accurate lung segmentor on chest x-rays using criss-cross attention and
  customized radiorealistic abnormalities generation,'' in [{\em Proceedings of
  The 2nd International Conference on Medical Imaging with Deep
  Learning}{\nolinebreak\hspace{0.1em}]},   {\bf 102},  457--467, PMLR (2019).

\bibitem{tang2019deep}
Tang, Y.-X., Tang, Y.-B., Han, M., Xiao, J., and Summers, R.~M., ``Deep
  adversarial one-class learning for normal and abnormal chest radiograph
  classification,'' in [{\em Medical Imaging 2019: Computer-Aided
  Diagnosis}{\nolinebreak\hspace{0.1em}]},   {\bf 10950},  1095018,
  International Society for Optics and Photonics (2019).

\bibitem{oh2018learning}
Oh, D.~Y. and Yun, I.~D., ``Learning bone suppression from dual energy chest
  x-rays using adversarial networks,'' {\em arXiv preprint arXiv:1811.02628}
  (2018).

\bibitem{yang2017cascade}
Yang, W., Chen, Y., Liu, Y., Zhong, L., Qin, G., Lu, Z., Feng, Q., and Chen,
  W., ``Cascade of multi-scale convolutional neural networks for bone
  suppression of chest radiographs in gradient domain,'' {\em Medical image
  analysis}~{\bf 35},  421--433 (2017).

\bibitem{eslami2019image}
Eslami, M., Tabarestani, S., Albarqouni, S., Adeli, E., Navab, N., and
  Adjouadi, M., ``Image to images translation for multi-task organ segmentation
  and bone suppression in chest x-ray radiography,'' {\em arXiv preprint
  arXiv:1906.10089}  (2019).

\bibitem{Goodfellow_GAN}
Goodfellow, I., Pouget-Abadie, J., Mirza, M., Xu, B., Warde-Farley, D., Ozair,
  S., Courville, A., and Bengio, Y., ``Generative adversarial nets,'' in [{\em
  Advances in Neural Information Processing
  Systems}{\nolinebreak\hspace{0.1em}]},  (2014).

\bibitem{tang2019ct}
Tang, Y.-B., Oh, S., Tang, Y.-X., Xiao, J., and Summers, R.~M.,
  ``{CT}-realistic data augmentation using generative adversarial network for
  robust lymph node segmentation,'' in [{\em Medical Imaging 2019:
  Computer-Aided Diagnosis}{\nolinebreak\hspace{0.1em}]},   {\bf 10950},
  109503V, International Society for Optics and Photonics (2019).

\bibitem{tang2019tuna}
Tang, Y., Tang, Y., Sandfort, V., Xiao, J., and Summers, R.~M., ``{TUNA-N}et:
  Task-oriented unsupervised adversarial network for disease recognition in
  cross-domain chest x-rays,'' in [{\em International Conference on Medical
  Image Computing and Computer-Assisted
  Intervention}{\nolinebreak\hspace{0.1em}]},   431--440, Springer (2019).

\bibitem{isola2017image}
Isola, P., Zhu, J.-Y., Zhou, T., and Efros, A.~A., ``Image-to-image translation
  with conditional adversarial networks,'' in [{\em Proceedings of the IEEE
  Conference on Computer Vision and Pattern
  Recognition}{\nolinebreak\hspace{0.1em}]},   1125--1134 (2017).

\bibitem{zhu2017unpaired}
Zhu, J.-Y., Park, T., Isola, P., and Efros, A.~A., ``Unpaired image-to-image
  translation using cycle-consistent adversarial networks,'' in [{\em
  Proceedings of the IEEE International Conference on Computer
  Vision}{\nolinebreak\hspace{0.1em}]},   2223--2232 (2017).

\bibitem{wang2004image}
Wang, Z., Bovik, A.~C., Sheikh, H.~R., Simoncelli, E.~P., et~al., ``Image
  quality assessment: from error visibility to structural similarity,'' {\em
  IEEE transactions on Image Processing}~{\bf 13}(4),  600--612 (2004).

\bibitem{seshadrinathan2010study}
Seshadrinathan, K., Soundararajan, R., Bovik, A.~C., and Cormack, L.~K.,
  ``Study of subjective and objective quality assessment of video,'' {\em IEEE
  Transactions on Image Processing}~{\bf 19}(6),  1427--1441 (2010).

\bibitem{shih2019augmenting}
Shih, G., Wu, C.~C., Halabi, S.~S., Kohli, M.~D., Prevedello, L.~M., Cook,
  T.~S., Sharma, A., Amorosa, J.~K., Arteaga, V., Galperin-Aizenberg, M.,
  et~al., ``Augmenting the national institutes of health chest radiograph
  dataset with expert annotations of possible pneumonia,'' {\em Radiology:
  Artificial Intelligence}~{\bf 1}(1),  e180041 (2019).

\bibitem{Simonyan15}
Simonyan, K. and Zisserman, A., ``Very deep convolutional networks for
  large-scale image recognition,'' in [{\em International Conference on
  Learning Representations}{\nolinebreak\hspace{0.1em}]},  (2015).

\end{thebibliography}
\bibliographystyle{spiebib} 

\end{document}